\def\year{2021}\relax
\documentclass[letterpaper]{article} 
\usepackage{aaai21}  
\usepackage{times}  
\usepackage{helvet} 
\usepackage{courier}  
\usepackage[hyphens]{url}  
\usepackage{graphicx} 
\usepackage{natbib}  
\usepackage{lipsum}
\usepackage{booktabs}
\usepackage{amsmath}
\usepackage{caption} 
\let\footnote=\endnote
\usepackage{etoolbox}
\makeatletter
\patchcmd{\@verbatim}
  {\verbatim@font}
  {\verbatim@font\tiny}
  {}{}
\makeatother

\title{Accurate and Interpretable Machine Learning for Transparent Pricing of Health Insurance Plans}
\author {
\parbox{\linewidth}{\centering
        Rohun Kshirsagar,
        Li-Yen Hsu,
        Vatshank Chaturvedi\textsuperscript{*},
        Charles H. Greenberg, 
        Matthew McClelland,
        Hilaf Hasson\textsuperscript{*},
        Anushadevi Mohan,
        Wideet Shende,
        Nicolas P. Tilmans,
        Renzo Frigato,
        Min Guo,
        Ankit Chheda,
        Meredith Trotter,
        Shonket Ray,
        Arnold Lee,
        Miguel Alvarado \\
        }
}
\affiliations{
  Lumiata Data Science and Engineering \\
  rohun@lumiata.com \\
  \textsuperscript{*}Work completed prior to joining Amazon
}

\begin{document}
\maketitle

\begin{abstract}
Health insurance companies cover half of the United States population through commercial employer-sponsored health plans and pay 1.2 trillion US dollars every year to cover medical expenses for their members.  The actuary and underwriter roles at a health insurance company serve to assess which risks to take on and how to price those risks to ensure profitability of the organization.  While Bayesian hierarchical models are the current standard in the industry to estimate risk, interest in machine learning as a way to improve upon these existing methods is increasing. Lumiata, a healthcare analytics company, ran a study with a large health insurance company in the United States.  We evaluated the ability of machine learning models to predict the per member per month cost of employer groups in their next renewal period, especially those groups who will cost less than 95\% of what an actuarial model predicts (groups with ``concession opportunities"). We developed a sequence of two models, an individual patient-level and an employer-group-level model, to predict the annual per member per month allowed amount for employer groups, based on a population of 14 million patients. Our models performed 20\% better than the insurance carrier's existing pricing model, and identified 84\% of the concession opportunities. This study demonstrates the application of a machine learning system to compute an accurate and fair price for health insurance products and analyzes how explainable machine learning models can exceed actuarial models' predictive accuracy while maintaining interpretability.
\end{abstract}
\section*{Introduction}
 The recent explosion of available electronic health record (EHR) and insurance claims data sets, coupled with the democratization of statistical learning algorithms, has set the stage for machine learning (ML) applications to fundamentally transform the healthcare industry. Employer-sponsored health insurance (ESI) currently covers 150 million Americans \cite{USChallengeHCC}. With numerous subsidies in place to make ESI more affordable \cite{EmpDropESIACA, TaxExclES}, it is by far the most popular option for obtaining health insurance in the United States \cite{CBOesi}. Since the passage of the Affordable Care Act (ACA) in 2009, premiums for single and family ESI plans have increased 50\% and deductibles have doubled, making affordability of care a major issue for many Americans.  Thirty four percent of patients on ESI plans are reportedly unable to pay for an unexpected bill of \$500 and over 50\% skip or postpone medical care and prescription fills due to cost \cite{EmpHealthBenefitsSurvey}. By making healthcare more affordable, it increases the chance patients will receive needed medical care and refill medications in a timely manner.  Increasing affordability improves patient health outcomes and quality of life, reducing familial strain of medical debt, postponing major household spending, and obviating the need to hold multiple jobs \cite{EmpHealthBenefitsSurvey}.

However, ESI premiums are the dominating source of revenue for US-based health insurance companies, many of which are among the Fortune 500 companies\footnotetext[1]{https://fortune.com/fortune500/2019/}. As such, accurate rate-setting is a crucial component of revenue and membership growth for insurance companies; setting inadequate rates can mean the difference between profitability and unprofitability \cite{RatAndUnder}. Traditionally, health insurance companies set rates using a combination of actuarial science and underwriting judgement.  Actuaries apply statistical modeling to claims data to set premiums for each employer-group; underwriters use the actuary's predicted rate alongside non-claims data (e.g. health questionnaires) to decide which groups to cover and what their rates will be. Accurate rate setting is essential to balance customer retention against business viability because the insurer needs to retain a group for several years before the account becomes profitable. Hence, insurers are willing to reduce rates in the near-term, in exchange for the chance of a longer term relationship (i.e. greater \textit{persistency}) \cite{PerGrpHealth}.  Good financial standing allows insurers to focus on growing their business, influencing patient health outcomes, improving customer experience, and increasing efficiency \cite{UnderExcel}. Reduced renewal premiums can align patients' financial interests and the insurer’s strategic interests. 

Within the ESI market, the $<$500 employer group segment (employer groups with fewer than 500 enrollees) is highly transactional, particularly during peak season (near January 1st of each year). The largest insurance carriers process tens of thousands of new  and renewal business quotes.  For instance, the carrier in this pilot study averaged 130 presale quotes and 70 renewal quotes per underwriter.  Some aspects unique to the $<$500 market are the: 
(1) Complex array of funding arrangements, including: fully-insured, level-funding Administrative Services Only (ASO, i.e. self-insured), pay-as-you-go and monthly shared-risk models\footnotetext[2]{Independence Blue Cross: Large Group Underwriting Guidelines} \cite{GrowEmpInt}
(2) high-risk, high-reward nature of this segment:  the $<$500 market comprises one third of medical insurance customers but yields one half of earnings, providing opportunities to yield higher profit margins, and
(3) lower persistency compared to larger clients (the average account length is 5 years for the $<$500 market vs $>$7 years for larger groups).

Due to the large volume of cases in the $<$ 500 market, underwriters could benefit from an additional highly-accurate signal to increase the efficiency of their work.  This signal would need to be easily interpretable and broadly applied for measurability. The unique circumstances of the $<$500 market provide an opportunity for improved renewal rate development, driving price optimization and increasing persistency in a risk-rich environment focused on strong, sustainable customer-base growth.

Although ML-based approaches on claims data are widely applied in clinical contexts (e.g. \citeauthor{DM2} \citeyear{DM2}; \citeauthor{CostBloom} \citeyear{CostBloom}; \citeauthor{scalable} \citeyear{scalable}), few studies incorporate ML approaches for underwriting group health insurance.  To our knowledge, there are no models of individual health risk used to form more competitive pricing for a group. Here, we show that ML approaches in medical underwriting can provide: (1)  improved accuracy and broader applicability, facilitated by modeling cost and risk at the individual and group levels; and (2) interpretability for non-experts to use.  


We jointly ran our study between Lumiata, a healthcare analytics company, and a large insurance company in the US, referred to here with the pseudonym ``Delphi".

\section*{Materials and Methods}


\subsection{Participants and Setting}
We drew from a 14 million member population representative of Delphi’s entire fully-insured employer-group customer base (i.e. their \textit{book of business}) from 2015 to 2017. We used this population to train and tune our models and then predicted the annual cost of groups in a separate ``holdout" set. The holdout set consisted of 648 employer groups (referred  to here simply as ``group") with renewal dates varying from 05/01/2016 to 04/01/2017. There were a total of 349,715 members still actively enrolled as of their respective group’s renewal date. Using the holdout set, we evaluated our model by its performance of predicted cost incurred during the 12-month period starting from their group's renewal date (the ``projection period").

\begin{figure}
 \centering
  \includegraphics[width=80mm, scale=0.4]{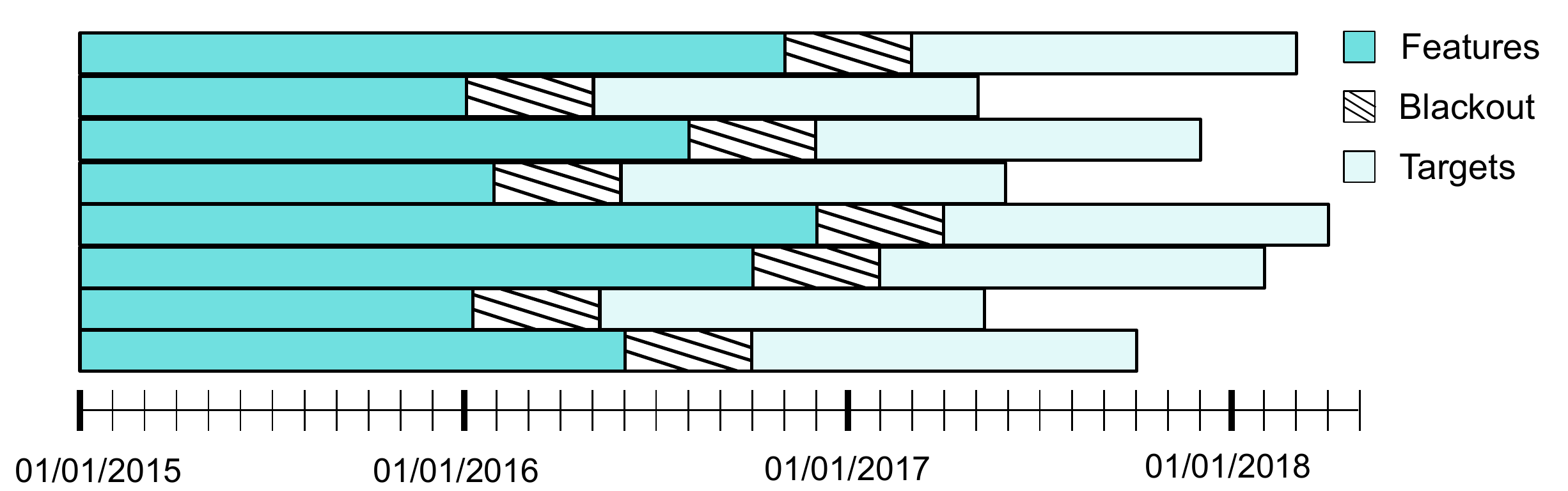}
  \caption{Holdout set ``dynamic time-slicing".  This picture shows the way the holdout set groups are time-sliced according to their renewal dates, and thus have different lengths of historical data (shown as ``Features") and different projection periods (shown as ``Targets") with respect to each other. Lumiata was blinded from the information in blackout and projection periods before submitting predictions to Delphi.}
  \label{fig:fig1}
\end{figure}

In the holdout set, Delphi censored the data recorded four months prior to the renewal date for each group (the ``blackout period"); group underwriting is usually done several months before the new contract year in order to create the renewal quotes presented to employer groups. The 12 months before the blackout period comprise the ``experience period", ending on the ``slice date". For example, if a group's renewal date was 05/01/2017, the experience period was the entire 2016 calendar year and 05/01/2017 to 04/30/2018 was the cost projection period (Figure~\ref{fig:fig1}). 

\begin{figure}
	\centering
  \includegraphics[width=70mm]{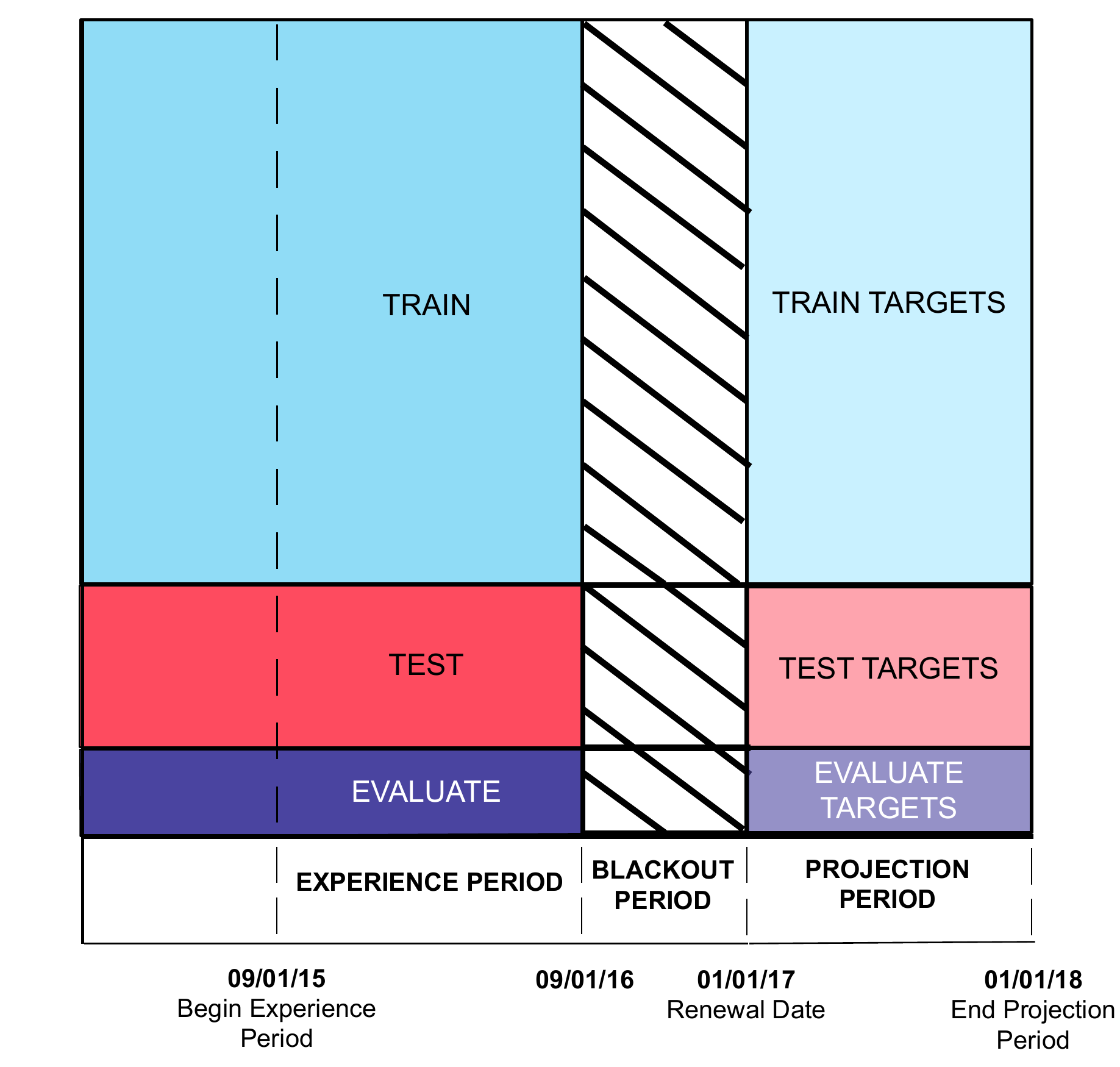}
  \caption{``Splitting" and ``time-slicing" for the training data. This figure shows how we split and sliced the data of groups and members enrolled on 08/31/2016 to train and validate our models. Groups were split using a 70:20:10 train:test:evaluate ratio.}
  \label{fig:fig2}
\end{figure}

For this pilot study, we sliced the training data to select groups and members in a similar, but simpler way. Instead of dynamically slicing the groups, we imposed a fixed renewal date of 01/01/2017 and used only the data recorded until 08/31/2016 for feature extraction.  Therefore, the members and groups for which we predicted cost were those eligible as of the 08/31/2016 slice date. After filtering the training data for eligibility,$\sim$ 7.4 million patients were enrolled as of the slice date. In order to fine-tune and evaluate models, we split these eligible groups and members into ``train", ``test", and ``evaluate" sets. The data were split using a 70:20:10 ratio (Figure~\ref{fig:fig2}).



\subsection{Data Sources}
We built our models using medical, capitation and pharmacy claims, and lab and eligibility tables for Delphi's patients and groups. The medical claims tables contained cost information, International Classification of Disease (ICD-9 and ICD-10) diagnostic codes, and Current Procedural Technology (CPT) procedure codes at the claim-level. The capitation tables contained only cost information. All tables reported each claim's care setting: inpatient, outpatient, ancillary, emergency, primary care, or specialty care. The pharmacy claims tables contained National Drug Code (NDC) medication codes and cost for each drug prescribed to a patient and the written and fill dates for the drug prescriptions. The cost associated to claims were given as ``allowed" amounts (the amount paid by the insurer plus the member’s cost share). Lab tables contained Logical Observation Identifiers Names and Codes (LOINC) lab test codes.  Eligibility tables captured each patient's health plans, enrollment time periods, and plan benefits.


\subsection{Models}\label{sec:a2}
\subsubsection{Actuarial Models} To compare model performance between Delphi's actuarial models and Lumiata's ML models, we followed the best practices of actuarial science.  Actuarial models estimate group cost on a per member per month (pmpm) basis. The normalization unit is called a ``member month", which is defined as one month of enrollment for one member. We used member months to normalize cost because predicted pmpm cost for a group translates to the monthly premium charged to each member in this group. Actuarial models estimate group-level cost, treating each member month of medical history as independent across members and within a member.  

The following example shows the utility and limitation of this perspective. Consider three hypothetical groups X, Y, and Z, each having 100 members and costing \$1 million during the 2017 calendar year:
\begin{itemize}
\itemsep0em 
\item  Group X: each member is enrolled for 10 months - the pmpm cost equals \$1 million/(10 months $\times$ 100 members) = \$1,000 pmpm
\item  Group Y: each member is enrolled for five months - the pmpm cost equals \$1 million/(5 months $\times$ 100 members) = \$2,000 pmpm
\item  Group Z: each member is enrolled for five months and one member costs \$900,000, while the rest of the members cost \$100,000 total - the pmpm cost equals \$1 million/(5 months $\times$ 100 members) = \$2,000 pmpm
\end{itemize}
As a result of the pmpm formulation, two groups with the same cost but different member months will have a different pmpm cost (Group X vs Y or Z). If the group cost is highly concentrated on one individual, versus being evenly distributed amongst the members in the group, it does not necessarily reflect in the pmpm cost, (Group Y vs Z).  In contrast, 
a member-level cost prediction model views Group Y and Z differently, therefore better modeling cost at the group level.

Actuarial predictive models used for quoting renewal business rely on dozens of rating factors (input variables) to build a predicted rate for a given group.  The factors rely on pre-computed demographic, medical trend, pharmacy trend and other actuarial coefficients derived from patients across a large (usually exogenous) population using regression-based methods. These rating factor ``priors" are then used to assemble a predicted trend value for a particular group (called the ``manual rate", $MR$).  Historical claims for the group are used to create the ``experience rate" ($ER$), which is then blended with the manual rate to create the final prediction. The proportion of blending between the two quantities is called ``credibility" ($c$, where $0 \leq c \leq 1$; \citeauthor{buhlcred} \citeyear{buhlcred}). Thus, the predicted total cost of a group can be expressed as:
\begin{equation}   
{\rm predicted~cost} = c \cdot ER + (1 - c)  \cdot MR
 \label{eq:eq1}
\end{equation}

This formula is a linear Bayesian hierarchical model, and is the optimal linear least-squares solution for estimating the annual pmpm cost of an employer group, called the Buhlmann-Straub method \cite{LecNoteCred}. This approach is the industry standard, underlying most models in production at insurance companies; applications include pricing, plan design, and reserve setting \cite{GrpInsur, CredGrpMed}. As a group becomes more credible, the actuarial model can rely more on the medical claims history of the group as an indicator of future expenses. In the absence of credibility, the safer bet is to rely on population-level cost estimations using only age and sex (the manual rate). Actuaries use a group's member months to parameterize credibility \cite{buhlcred, CMSCred}.  A larger group enrolled for a shorter period of time (e.g. 1000 members $\times$ 6 months = 6000 member months) can have the same number of member months as a smaller group enrolled for a longer period of time (e.g. 200 members $\times$ 30 months = 6000 member months), making them equally credible.

The two equations below are examples of the type of experience rating and manual rating models used by Delphi:
\setlength{\belowdisplayskip}{0pt} \setlength{\belowdisplayshortskip}{0pt}
\setlength{\abovedisplayskip}{0pt} \setlength{\abovedisplayshortskip}{0pt}
\begin{multline*}
ER =  (TC - TSC ) (1 +  AT )^\frac{m}{12} \cdot  x_{\rm m} x_{\rm b} x_{\rm d}  +  \\  n_{\rm s}x_{\rm p} 
           + BC_{\rm p} (1 + AT_{\rm L} )^\frac{m}{12} \cdot x_{\rm ph} x_{\rm gp} x_{\rm dp} x_{\rm ip} \cdot mm
\end{multline*}
\\
\begin{multline*}   
MR = [ BC_{\rm med} (1 + AT_{\rm med})^\frac{m}{12} \cdot x_{\rm gm}x_{\rm dm}x_{\rm im}x_{\rm udm} + \\ BC_{\rm cap} (1 + AT_{\rm med})^\frac{m}{12}+ \\
            BC_{\rm ph}  (1 + AT_{\rm ph}  )^\frac{m}{12}  \cdot x_{\rm gph}x_{\rm dph}x_{\rm iph}x_{\rm udph} ] \cdot mm \\
\end{multline*}  
The definitions of independent variables in these two equations are shown in Table~\ref{tab:ind_var_desc}. For example, $x_{d}$ is the average of the group's member's age and sex factors, which are legal and widely used for large group underwriting in the United States.  The experience rate is linear in terms of the total claims ($TC$), and is combined with the manual rate and the group's member months (Eqn. \ref{eq:eq1}).

\begin{table}
 \small
 \begin{tabular}{|l|p{3cm}|p{3cm}|} 
 \hline 
 \textbf{Variable} & \textbf{Definition} & \textbf{Comment}\\ [0.5ex] 
\hline
 
 $AT$ & annual trend & medical and pharmacy trends combined \\ 
 
 $AT_{\rm L}$ & leveraged annual trend &  adjusted for the effects of pooling-point leveraging \\ 
 
 $AT_{\rm med}$ & medical annual trend & \\

 $AT_{\rm ph}$  & pharmacy annual trend  &\\

 $BC_{\rm cap}$ &  base capitation claims & pmpm cost of medical claims during the experience period\\ 

 $BC_{\rm med}$ &  base medical claims & pmpm cost of capitation claims during the experience period\\ 

 $BC_{\rm p}$ &  base pooled claims &  pmpm cost of pooled claims during the experience period\\ 

 $BC_{\rm ph}$ & base pharmacy claims & pmpm cost of pharmacy claims during the experience period\\ 

 $TC$ &  total claims & total cost during the experience period \\
 
 $TSC$ &  total shock claims & total cost of claims over the pooling level during the experience period \\
 
 $m$ & midpoint months & number of months between the midpoints of experience and projection periods  \\ 

 $mm$ &  member months & member months during the experience period\\
 
 $n_\text{s}$ &  number of shock claims & number of claims over the pooling level\\
$x_\text{b}$ &  benefit & \\ 

 $x_\text{d}$ &  demographic & based on the experience period\\ 

  $x_\text{dm}$  & demographic - medical & based on census\\

  $x_\text{dp}$ & demographic - pooling & based on census\\

  $x_\text{dph}$ & demographic - pharmacy & based on census\\

 $x_\text{gm}$ &  geographic area - medical &based on census \\ 

$x_\text{gp}$ & geographic area - pooling &based on census \\

  $x_\text{gph}$ &  geographic area - pharmacy & based on census \\

$x_\text{im}$ &  industry - medical & \\
 
  $x_\text{ip}$ & industry - pooling & \\

 $x_\text{iph}$ &  industry - pharmacy & \\

 $x_\text{m}$ &  maturation & \\ 
 
  $x_\text{p}$ &  pooling level & the cost threshold defining a shock claim (typically \$100,000) \\

 $x_\text{ph}$ &  pharmacy load & \\

 $x_\text{udm}$ &  utilization dampening - medical &  $1.2e^{-0.8{S}}$, where $S$ is medical cost share\\

 $x_\text{udph}$ &  utilization dampening - pharmacy & determined from a table using medical cost share \\

 \hline
\end{tabular}
\caption{Definitions of all the independent variables in the manual rate and experience rate equations. The most recent census was used for lookups. All $x$ variables are factors.}
\label{tab:ind_var_desc}
\end{table}

To improve accuracy, we modeled cost at both the individual and group levels.  We used a sequence of two models: (1) Individual-level model – predicting per month cost for a given member and  (2) Group-level model – predicting per member per month (pmpm) cost for a given group. Our approach contrasts with traditional actuarial methods which are heavily focused on group-level cost and lack individual-level information within each group. 
\subsubsection{Feature Engineering}\label{sec:a1}
We reshaped the claims and eligibility tables into longitudinal patient records per our proprietary data format, the ``Lumiata Data Model" (LDM).  From the LDM, we created member-level features (input variables) using information before the blackout period, based on techniques from the literature (e.g. \citeauthor{DM2} \citeyear{DM2}; \citeauthor{CostBloom} \citeyear{CostBloom}). Our demographic features were age and sex; all other features were time-dependent. Our time windows to compute a variety of features (e.g. diagnosis, medication, procedure, lab, and revenue codes, and cost and coverage) were: ``last three months", ``last six months", ``last one year", and ``anytime" prior to the blackout date.

In addition to ICD-9, ICD-10, CPT, NDC, and LOINC codes, we transformed the 
codes into their grouped counterparts based on organ-type (SNOMED), condition categories 
(HCUP, CMS-HCC and HHS-HCC), drug molecule (RxNorm and ATC), and our proprietary clinical grouper (Lumiata disease code). We 
derived features from these codes by calculating the log count of every unique code in each time window, and the summary statistics of each system (total count, unique code count, minimum count, maximum count, mean count, etc) in the ``anytime" window. The presence of revenue codes (binary) was computed in the ``anytime" window.

We computed features using their observed lab interpretations or values from the LOINC codes. These include (1) log counts of interpretations, i.e. ``high", ``low", ``abnormal", and ``normal", (2) whether the value was increasing, decreasing or flat across time points for the same test (one-hot encoded), and (3) if the interpretation was fluctuating across time (binary). For (2), the calculation was based on a t-test's $p$-value of a simple linear regression's slope.

Cost features are the most powerful features to predict future cost. In addition to the cumulative allowed cost in different time windows, we computed cost attributed to different care settings. The length of coverage was computed for all time windows. In total, we constructed more than 5 million possible features per patient. The resulting feature matrix was very sparse; most of the columns had no values at all. We reduced the dimension of the feature space 
before training a model using feature selection techniques discussed below.

\subsubsection{Individual-Level Model}
We regressed our first model on the allowed amount per month during the projection period for each member. We trained a gradient boosting tree that optimized the mean squared error (MSE) using the LightGBM package \cite{LightGBM} in the Python programming language \cite{python}. We chose LightGBM over linear regression and XGBoost due to its better performance.  To speed up training time and reduce over-fitting, we tested a variety of feature prevalence thresholds to reduce the features set to $\leq$100,000. We defined the prevalence of a feature as the fraction of non-zero values for this feature across patients in the training set. The model implicitly selected useful features to split on; the typical number of features was $\sim$ 4000 (see Table~\ref{tab:ind_feat}).

We ran the model recursively, using a different threshold (thus a different number of features) in each iteration. Evaluating on the test set reached a performance plateau at a threshold of $\sim$ 0.001, which we used for all subsequent models. We used the test set data for hyper-parameter tuning and early-stopping. Using the trained model, we made cost predictions for all the individuals in the train, test, evaluate, and holdout sets. 

Members enrolled as of 08/31/2016 were included in our train, test, and evaluate sets. Members who dropped out during the blackout period were filtered out before training; we did not train on a member's data if this member was not enrolled in a group on 01/01/2017. We only omitted these members during training, not during inference.





 






\begin{table}[t]
 \centering
 \begin{tabular}{l l l} 
 \hline
 \hline
feature type & value type & count   \\  
 \hline

age & numeric & 1  \\
gender & binary & 1 \\
cost & numeric & 28  \\
coverage & numeric & 4 \\
revenue code & binary & 70 \\

ATC log count & numeric & 31 \\
ATC summary stats & numeric & 7 \\

CMS-HCC log count & numeric & 11 \\

CPT log count & numeric & 771 \\
CPT summary stats & numeric & 6 \\
 
HCUP log count & numeric & 337 \\
HCUP summary stats & numeric & 7 \\  

HHS-HCC log count & numeric & 6 \\

ICD10 log count & numeric & 40 \\
ICD10 summary stats & numeric & 7 \\
ICD9 log count & numeric & 12 \\
ICD9 summary stats & numeric & 6 \\
LOINC log count & numeric & 115 \\
LOINC summary stats & numeric & 7 \\
LOINC interpretation log count & numeric & 117 \\
LOINC interpretation fluctuation & binary & 4 \\
LOINC value trend & binary & 88 \\

NDC summary stats & numeric &4 \\
RxNorm log count & numeric & 230 \\
RxNorm summary stats & numeric & 6 \\
SNOMED log count & numeric & 2082 \\
SNOMED summary stats & numeric & 7 \\

Lumiata disease code log count & numeric & 91 \\  

 \hline
 \hline
\end{tabular}

\caption{A summary of features used by the individual-level model. Note that only the ``anytime" window is used for revenue code and all 
the summary statistics. For LOINC value trend, each code corresponds to three one-hot encoded binary features, i.e., ``increasing", ``decreasing", ``flat",  
given a time window.}
\label{tab:ind_feat}
\end{table}

\subsubsection{Aggregation}
Our individual model predicts the per month cost in the projection period for the members who were enrolled in groups at the end of experience period. For our train, test, and evaluate sets, this date was 08/31/2016. For the holdout set, the slice date depended on the group’s renewal date. We then aggregated these predictions based on the members active on this date to obtain the mean of member-level cost predictions for each group. This quantity has a unit of pmpm and became the input of our group-level model described below.

\subsubsection{Group-Level Adjustment}
The aggregated mean prediction of a group can be thought of as predicted pmpm cost. However, this quantity only considers the 
members enrolled at the end of the experience period and assumes enrollment remains constant throughout the projection period. In reality, a group can grow or shrink during the blackout and projection periods, affecting the true pmpm. We regressed a second model on the true group-level pmpm cost to 
adjust the aggregated predictions. We experimented with different group-level features in this model and use the following (per group):
(i) mean cost of member-level predictions, (ii) mean member age, 
(iii) total number of member months for the group during the experience period, (iv) ``growth" feature, defined as the change in the number of members during the experience period divided by the total number of member months, (v) average length of member coverage,
(vi) fraction of experience period costs that were incurred during the final four months, before the blackout period (vii) fraction of high-cost members, defined as someone whose cost falls within the top 10\% of all members in the training set.
We then trained a LightGBM model that optimized pmpm Mean Absolute Error (MAE) and used the test set to perform hyper-parameter tuning and early-stopping. The mean of member-level cost predictions for each group highly correlates with the overall target pmpm cost for the group. With the additional features, the group-level model improved pmpm MAE by $\sim$ 10\% compared to the individual-level model alone.  

Note that a regression model optimizing MSE estimates the true mean of $y$ at a given $x$, while optimizing MAE is equivalent to estimating the true median. Thus, in order to make the aggregated individual-level predictions unbiased estimators for the group pmpm costs, we must optimize MSE instead of MAE at the individual level, even though we are pursuing MAE at the group level. 

Similar to the individual model, we trained the group model on groups still active during the projection period because only the groups that remained active will be included when evaluating the results. This operation was only done in model training and was not performed when doing inference using the trained model.




\subsection{Model Evaluation}


\subsubsection{ML Metrics Evaluation}
We adapted the standard metrics R-squared ($R^{2}$), MAE, and Gini index \cite{gini} to predicted pmpm cost; evaluation metrics are measured by comparing ``true pmpm cost" and ``predicted pmpm cost". Lumiata received claims data at the allowed amount level to preserve Delphi's pharmaceutical and provider reimbursement rates, whereas Delphi's production models predict the ``paid amount" for employer groups. As a result, Delphi and Lumiata modeled two different types of cost, which are not directly comparable because the allowed amount is usually greater than the paid amount (though they are highly correlated; see \citeauthor{AccRisk} \citeyear{AccRisk}). For business purposes, to make the predictions comparable, both teams agreed on two solutions. First, we computed ``normalized'' pmpm MAE, which equals pmpm MAE divided by the ``global pmpm cost". The global pmpm cost is the total (allowed or paid) amount divided by the total number of member months across all groups.  Second, we computed the “trend” versions of our predictions. The ``allowed trend" was defined as the pmpm allowed amount in the projection period divided by the pmpm allowed amount in the experience period; the ``paid trend" was defined similarly.  Trend calculations are ubiquitous in actuarial science \cite{SOATrend}.  ML and actuarial metrics used in this context provide strong directional support for one model over another, as we will see.

\subsubsection{Lift Plot and Concession Opportunities}
While ML performance metrics like MAE and $R^{2}$ are ubiquitous in the tech industry, they often lack direct connection to concrete business-level Key Performance Indicators (KPIs) in other fields \cite{AnaTransl}. To address this gap, we computed a KPI called a ``lift plot", illustrating the implications of Lumiata's model in a production context. The lift plot is made using the following steps: (1) compute Lumiata's predicted allowed trend divided by Delphi's predicted paid trend (the ``trend ratio"),
(2) rank groups according to the trend ratio,
(3) segment the ranked groups into deciles,
(4) compute the actual-to-expected (A/E) ratios within each decile, where ``A" is the true paid amount of a given decile, and ``E" is Delphi's predicted paid amount of the same decile,
(5) normalize the A/E of all the ten deciles by the global A/E of the holdout set, and
(6) plot the ``oracle" model (a model that perfectly predicts pmpm allowed trend).
The predicted paid amount of a given decile is the sum of the predicted pmpm paid amount times the projected member months for all groups in this decile.  The projected member months of the projection period is 12 months times the number of members at the end of the experience period.

A group is considered to be a ``concession opportunity at the 5\% level" if the group's rates can be reduced by 5\% while maintaining profitability (A/E remains $<$1). In other words, the true paid trend ratio (true pmpm paid trend divided by Delphi’s predicted pmpm paid trend) is $<$ 0.95.  We used precision and recall to evaluate concession opportunity identification performance \cite{PrecRecall}. In order to identify concession opportunities at the 5\% level, we applied a decision rule of $<$ 1 using a predicted ``trend ratio" (Lumiata's predicted pmpm allowed trend divided by Delphi’s predicted pmpm paid trend). That is, if Lumiata's predicted pmpm allowed trend was lower than Delphi’s predicted pmpm paid trend for a group, we asserted that group was a concession opportunity at the 5\% level. We used a decision rule of $<$ 1 instead of $<$ 0.95 in order to achieve a higher recall.


\section*{Results}
\subsubsection{Model Performance}\label{sec:performance}
Lumiata's model was 20\% better in normalized pmpm MAE, 26\% better in pmpm $R^2$ than Delphi’s model, and 2\% lower in Gini index than Delphi's model (Table \ref{tab:2}). Lumiata correctly identified 84\% of the groups in the holdout set that had concession opportunities at the 5\% level.


\begin{table} 
 \centering
 \begin{tabular}{c c c c} 
 \hline \hline
 model & MAE & $R^2$ & Gini index\\ 
 \hline
Delphi  & 0.239 & 0.265 & \textbf{0.600}\\ 
Lumiata & \textbf{0.192} & \textbf{0.334} & 0.588 \\
 \hline \hline
\end{tabular}
\caption{Performance metric results for Lumiata vs Delphi on a normalized pmpm basis.  Bold indicates higher performance for that metric.}
\label{tab:2}
\end{table}

Lumiata's predicted pmpm allowed trend had a 65\% precision and 84\% recall in identifying concession opportunities at the 5\% level, and a 56\% precision and 85\% recall in identifying concession opportunities at the 10\% level. For comparison, an ``oracle" model had an 84\% precision and 96\% recall at the 5\% level, and a 73\% precision and 97\% recall at the 10\% level. This similarity in precision and recall indicates Lumiata's model was near optimal for predicting concession opportunities at the 5\% level. 
\subsubsection{Practical Application} 
\begin{figure}[t]\begin{center}
  \includegraphics[width=80mm]{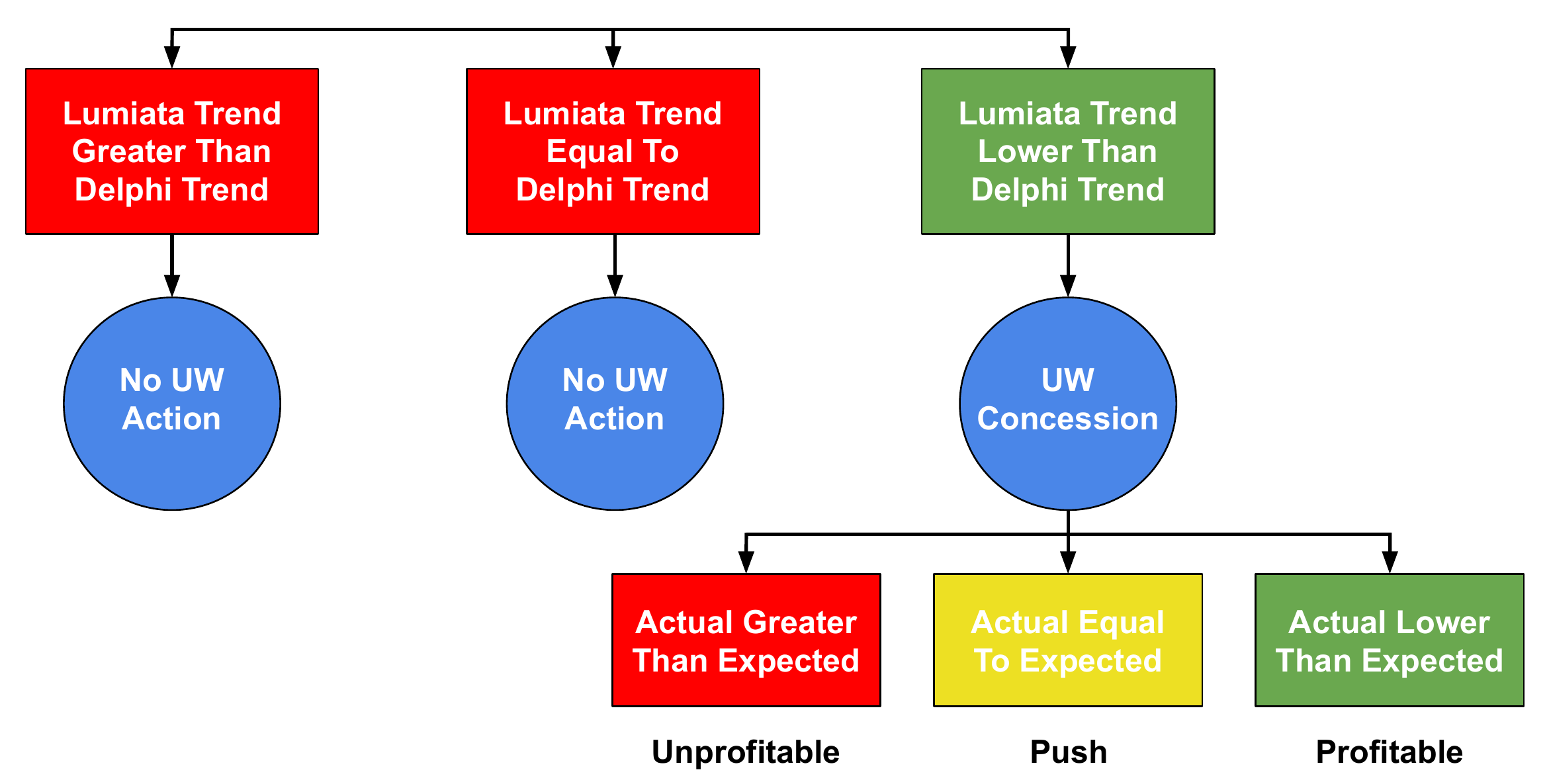}
  \caption{Flow-chart of the ``stop-light" process implementing Lumiata’s model output. ``UW" stands for underwriting.
}
  \label{fig:stoplight}
  \end{center}
\end{figure}

Operationalizing Lumiata's model relied on the ``stop-light principle" to make the model output interpretable to non-data scientists. The decision process derived from Lumiata's model is: (1) If Lumiata's predicted trend is less than Delphi's predicted trend, then the model suggests an underwriter give a concession of at least 5\% on that group's renewal quote (Green).  
(2) If Lumiata's predicted trend is equal to or greater than Delphi's predicted trend, no action is taken (Yellow or Red, see Figure \ref{fig:stoplight}).

\begin{figure}\begin{center}
  \includegraphics[width=80mm]{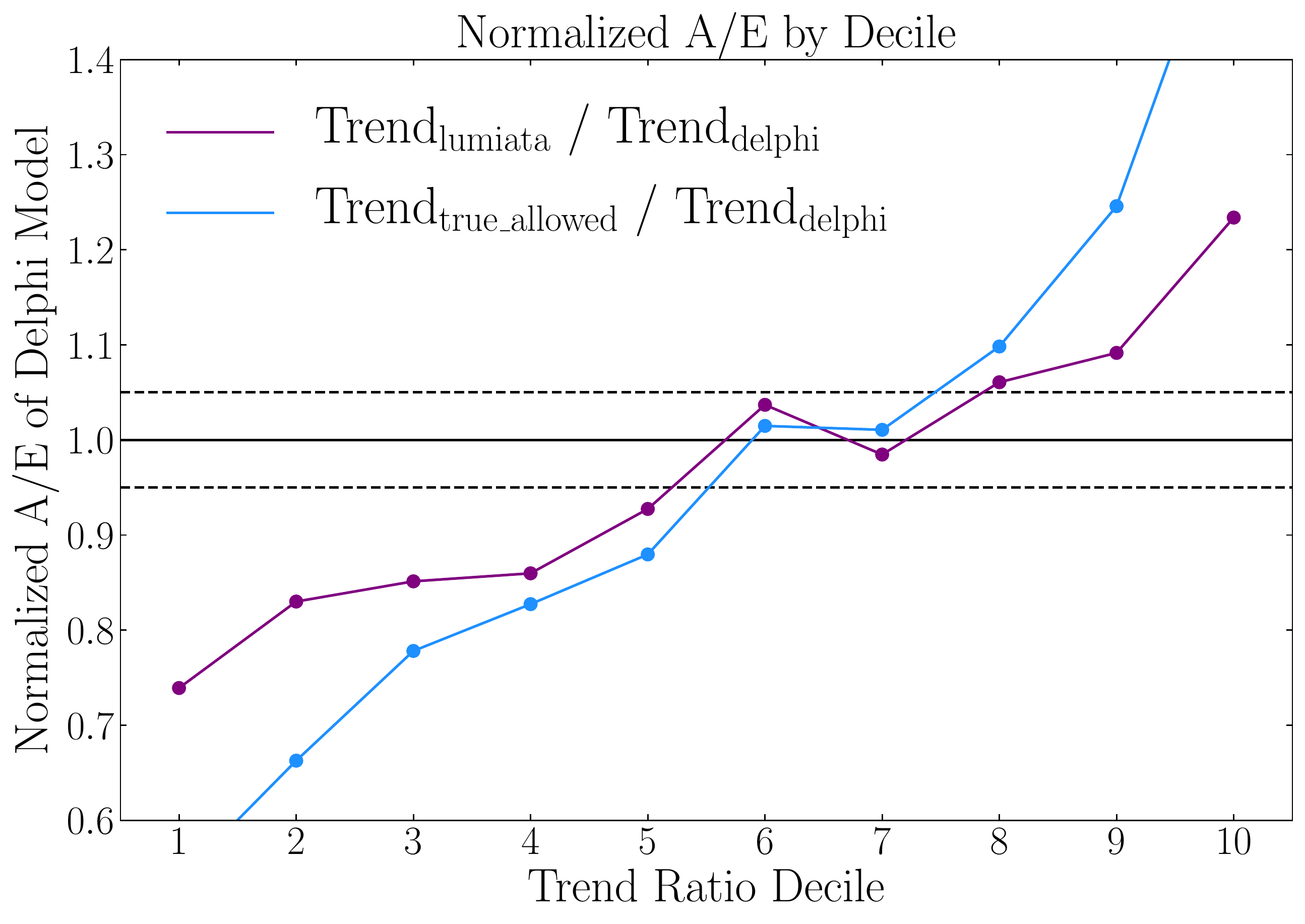}
  \caption{Lift plot for the holdout set using Lumiata (purple) and true allowed (blue) trend ratios. The trend ratio of a model is defined as its predicted trend divided by Delphi's predicted trend. The solid black line indicates A/E = 1.  The dotted black lines indicate where the A/E = 1.05 or 0.95.  Concession opportunity groups are in deciles below the 0.95 dotted line.}
  \label{fig:lift}
  \end{center}
\end{figure}

The lift plot in Figure~\ref{fig:lift} shows the result of using Lumiata's predicted pmpm allowed trend in this decision process. The average A/E per decile of Delphi's model varies with the decile of Lumiata trend ratio (Lumiata's predicted allowed trend divided by Delphi's predicted paid trend) and true allowed trend ratio (true allowed trend divided by Delphi's predicted paid trend), respectively. The A/E of the bottom five Lumiata trend ratio deciles are below 0.95, meaning Lumiata's trend ratio can select groups for a rate drop of $\ge$ 5\% with good accuracy. The median trend ratio was ~0.89 which yielded higher precision than a decision rule of 1.0.

Had Delphi implemented our model for this pricing period, their underwriters could have dropped the renewal quote 5\% or more for approximately half of the groups while retaining profitability. An ``oracle" model (blue line in Figure~\ref{fig:lift}) identified the same number of decile concession opportunities at the 5\% level as our model (five deciles). In this application, our model acts as a decision-support tool to give efficient guidance to underwriters.

\section*{Deployment}\label{sec:deployment}
The goals of Lumiata's pilot study with Delphi were to prove Lumiata's (1) ML could improve over an established industry methodology, and (2) tech stack could deliver monthly predictions for Delphi's renewal business groups. We designed a Kaggle-style competition\footnotetext[2]{https://www.kaggle.com/} between the two companies, with two holdout sets - a preliminary holdout set (658 groups) and a final holdout set (which we call ``holdout set" throughout this paper - 648 groups). We had one chance to compare our predictions to Delphi's on the final holdout set; the success or failure of the pilot study was predicated on whose model had the best group-level cost predictions.  
\subsubsection{Quality Assurance}
Due to several delicate calculations needed to assemble the predicted/true allowed trend predictions, we computed non-prediction fields to rule out non-data science confounding factors before running the results analysis. These fields included (per group):
(i) number of members enrolled at the end of experience period,
(ii) number of member months in the experience period,
(iii) true allowed amount in the experience period, and
(iv) predicted allowed amount in the projection period.
After receiving the censored information, we computed the:
(i) number of members at the beginning of projection period,
(ii) number of member months in the projection period, and
(iii) true allowed amount in the projection period.
Due to a strong data model built off of FHIR\footnotetext[3]{https://wiki.hl7.org/FHIR} and solid compute infrastructure, we were able to iterate and fix bugs quickly, until our calculations of the experience period non-prediction columns matched Delphi's within 5\%. Completion of this analysis allowed quick and self-evident comparison of Lumiata’s and Delphi’s models’ performance metrics. Additionally, the non-prediction fields found their way into roll-out plan below.

\subsubsection{Data Challenges}
For model comparison, we had to address discrepancies between Lumiata's and Delphi's claims data sets.  Two major differences in the claims data were: 
(1) Delphi’s models used paid amount and Lumiata used allowed amount, and
(2) Delphi used the ``paid date", but requested Lumiata use the ``encounter date" for allowed amounts (Delphi felt ``encounter date" was more appropriate for patient-level cost predictions). 
To solve these issues, we computed the ``allowed trend" compared to the ``paid trend" (Figure \ref{fig:lift}).  Calculating the ``allowed trend" required we compute three other quantities per group (``number of members at the end of experience period", ``member months in the experience period", and ``allowed amount in the experience period"), and the predicted allowed amount for the group.  Accurately computing these quantities was more difficult than expected due to the consequences of dual paid/allowed conventions and time-dependent patient enrollment:

(i) Calculating the ``allowed amount in the experience period" differs depending on whether ``paid date" or ``encounter date" is used. Using the paid vs encounter date segmented claims differently into a group’s experience, blackout, and projection periods. For example, some claims were denied or paid claims could be reversed.  Hence, claims data filtered on encounter date in the holdout set experience period were sometimes absent from the unblinded holdout set's \textit{experience} period.

(ii)  The number of members enrolled in a group at the end of the experience period could change during the blackout and projection periods.  For example, members can shift between different groups due to job or spousal health plan changes. Also, enrollment was updated on the 15th of the month, but we calculated ``number of members at end of experience period" and ``member months in the experience period" for the first of each month, making our enrollment data two weeks out of date.

\subsubsection{Roll-out Strategy}\label{sec:deployment}
Delphi requested Lumiata provide monthly group-level cost predictions and concession opportunities for groups up for renewal within the next four months (Delphi renews groups throughout the year).  Delphi asked for a seven day cost-prediction turn around time and fixed model feature set each quarter.  In response, we built a platform that can create LDMs for 1 TB of claims data in under an hour and produce highly optimized group-level cost predictions in under four hours.  

To streamline deployment, we proposed the following plan to Delphi:
(1a) Each month, Delphi sends a moving three year snapshot of their claims and eligibility tables.  
(1b) Delphi sends their non-prediction field calculations and their paid trend group cost predictions for groups up for renewal.
(2) Lumiata updates the existing patients’ LDMs and adds new patients from the eligibility file.
(3) Lumiata updates the group cost prediction model, training using the prior 12 months as the projection period, four months prior as the blackout, and prior 14 months as the experience period.
(4) Lumiata creates feature vectors including the new projection/blackout period claims information.
(5) Lumiata applies the updated member- and group-level models to the claims data to produce group-level allowed trend predictions.
(6) Using Delphi's paid trend predictions from 1b and the ``stop-light" principle (Figure \ref{fig:stoplight}), Lumiata recommends whether to drop the rate by 5\% for each group.
(7) Lumiata sends its non-prediction attributes and the allowed trend predictions to Delphi with one recommendation per group up for renewal. All non-prediction fields must agree within $\leq$ 5\% between Delphi and Lumiata's calculations.
(8) Delphi verifies receipt of the data and the results are consumed by their actuarial and underwriting teams. 
All files are transferred to and from Lumiata's platform (hosted on Google Cloud Platform) using sFTP.

\subsubsection{Transparent Rate Setting}\label{sec:deployment}
Actuarial models have an essential property: they are ``explainable" because a prediction can be decomposed into discrete multiplicative factors with an inherent interpretation.  For example, ``geographic area factor" = 0.9 means people in a particular zip code cost 10\% less than the mean, so the base rate is adjusted (multiplied) by 0.9 for members from this zip code.  This degree of explainability is crucial because actuaries need to file rates annually for individual/small group markets with the state insurance commissioner to ensure the factors used to produce the rate are compliant with legal guidelines.  Furthermore, an underwriter needs to be able to explain how she arrived at a particular rate to a customer.
A critique of ML models is that they lack explainability in terms of what input variables may have contributed to a particular prediction \cite{xai}.  However, explainable ML in healthcare is must-have, touching upon fundamental issues of bias, transparency, and reasonableness of ML model predictions.

Shapely values, a game theoretic algorithm, enable common ML algorithms to output feature weights specific to a prediction \cite{Shap}.   We applied the Python package SHAP\footnotetext[4]{https://github.com/slundberg/shap} to our LightGBM models to yield member-level explanations.  According to the algorithm, a model $f$ and a member feature vector $x$ admit an additive decomposition in terms of the mean value of the model $\mathbf{E}[f(x)]$ and SHAP values $\phi_{i}$ (interpreted as dollar value pmpm amounts) for the $i$th feature.  For our member- and group-level models, $f_{1}$ and $f_{2}$ are:
\begin{equation}   
{f_{1}(x)} = \mathbf{E}[f_{1}(x)] + \sum_{i=1}^{3996} \phi_{i, 1}
 \label{eq:eq2}
\end{equation}
\begin{equation}   
{f_{2}(x)} = \mathbf{E}[f_{2}(x)] + \sum_{i=1}^{7} \phi_{i, 2}
 \label{eq:eq3}
\end{equation}
Given our member and group model sequence, our group model (\ref{eq:eq3}) admits an additive decomposition in terms of member-level $\phi_{i, 1}$ .  The mechanics in (\ref{eq:eq3}) are similar to an actuarial formula, in that SHAP values for a group's members' features additively “adjust” the mean pmpm cost of the model over all members and groups, while the actuarial factors for a particular group multiplicatively “adjust” the mean pmpm cost of their entire patient population.  We often found that $\leq$500 features' SHAP values account for 95\% of a cost prediction, for each group.  However, the specific features involved varied by group.

The transparency afforded by the group-level SHAP values provides the opportunity to explain a rate adjustment to a customer in dollar pmpm amounts using the specific risk drivers for that group and modify a rate given by the ML model by the expected change in cost for specific drugs and services.  For instance, if the price of Glipizide, a drug to treat type-2 diabetes, will drop for an insurer next year by 20\%, imagine a world where the insurer can multiply the SHAP values corresponding to Glipizide-related pharmacy costs by 0.80 for all the members on Glipizide, thus lowering the projected rate.  These mechanics would be similar to current actuarial methods, making them easy to implement. Furthermore, greater model transparency could increase patient adherence to prescribed medications. As drug prices rise and more patients purchase high-deductible plans, patients have to pay higher out of pocket costs for drug treatment, and patient medication adherence declines \cite{DrugCost}. Insurers could use the SHAP values from the patient-level model to identify drugs driving up projected cost for that group, and suggest the prescribing doctor offer a lower cost alternative drug with similar efficacy. This provides a win-win opportunity, lowering drug cost for the payer, and improving patient adherence to the cheaper drug through increased affordability. 


\section*{Discussion}
Here, we demonstrate that: (1) ML approaches can significantly improve the accuracy and efficiency of group health insurance underwriting and (2) ML models can offer comparable interpretability to traditional actuarial methods. Our contributions provide clear direction for how to improve the efficiency and predictive performance of underwriting for employer-based insurance and how to lower the cost for members in groups of any size. Our ML-based approach improved MAE over actuarial models across the book of business: $>$500-member groups showed an improved performance. 

Our model shows the most improvement over actuarial models in situations where the group size is $\leq$500 and/or the group claims experience is relatively short ($<$8 months).  In these situations, the groups are not yet credible, so actuarial models perform sub-optimally. We believe the success of our model was due to modeling costs: (1) at the individual level; by contrast, actuarial methods aggregate medical history across group members, (2) with models that perform well with skewed distributions; the cost of care in an insurance population is often gamma distributed, making the ML method of gradient boosting trees, like LightGBM, highly effective \cite{gradboost}, (3) using a model-agnostic approach to select features relevant for predictions, and (4) by combining individual- and group-level models to produce the final predictions.

We treated all members in the training set as if their group's renewal was 01/01/2017, with a four month blackout period starting on 08/31/2016. However, patients often try to ``fit in" healthcare services before their next renewal period because they are likely near or above their plan’s deductible and hence will have these services fully covered by the insurance company. When patient records were ``sliced" uniformly on 08/31/2016, there was a chance that this useful information would be lost and negatively impact the model performance. We found that dynamically ``slicing" patient records according to their group’s renewal date (instead of all patients on 08/31/2016), and training the models with this feature setup, the overall results on the holdout set remained roughly the same in terms of MAE and $R^2$. Model performance likely didn't improve because claims data are inherently fuzzy with dates; claims can take variable amounts of time to get paid, and hence a model that tries to learn precise date information will add some but not meaningful value. To account for fuzzy date information, we aggregated features over 3 months, 6 months, etc (see Methods).

As ML models are increasingly compared to more traditional statistical techniques, the most appropriate study design and model evaluation metrics should be examined. For example, the holdout set data was ``out of sample" (i.e. using patients/groups unseen before by the model) but not ``out of time" (i.e. projecting costs for time periods subsequent to 2017).  Furthermore, claims data are not time-stationary (e.g. new drugs and treatments will be developed), so the expected model performance may not be perfectly realized in practice, but the relative difference between the models should hold. Also, we obtained a slightly worse Gini index than Delphi, despite our much better $R^2$, MAE, and lift plot (Table \ref{tab:2} and Figure \ref{fig:lift}).  This discrepancy can occur because the Gini index is a ranking-based metric, whereas a regression model minimizes the prediction errors. One difficulty is the Gini index is not a differentiable quantity. Future work should develop algorithms to address this problem.  

Data quality was crucial to our success. Alignment on non-prediction fields between Lumiata and Delphi ruled out errors in the data, pipeline, or output, improving communication across teams and increasing efficiency. These calculations must be automated for developing and productionalizing a medical underwriting ML application, due to the large size of data sets and rapid turnaround of results.

Due to its highly applied nature, some operational realities limit our study's evaluation. A challenge for validating our predictions is the long feedback cycle (20 months). Also, not all of Lumiata's concession recommendations could be granted due to a variety of quantitative and judgement factors under the underwriter and insurer's discretion. 

Additionally, we could not determine if our individual-level model was racially biased, because we did not receive patient ethnicity data.  Avoiding racial bias is important as previous studies have found evidence of racial bias in commercial cost prediction models used for clinical management \cite{RacialBias}.  Historically, poorer minorities under-utilized healthcare services due to mistrust of the system and confusion about how to navigate it \cite{RacialBias}. However, because our response variable is not a clinical outcome but a financial one, we think this effect on pricing may be less significant.  More work will be needed to better understand the effect of pricing insurance more affordably for minority patients, predicated on their less frequent utilization of the healthcare system.

In practice, ML approaches can help insurers be more competitive, avoiding adverse risk. It can result in the design of more ``exotic" funding arrangements due to better predictive power of patient health, following the industry trend towards capitated payments \cite{mngdcare}.

Unlike previous ML models in healthcare \cite{xai}, our model output is interpretable by a non-technical user, simplifying operationalization (Figure~\ref{fig:stoplight}). A user does not need to understand the inner workings of our algorithm to apply our output as a multiplicative adjustment factor to their existing actuarial models and can output the most important group-specific risk factors.

\section*{Conclusions}
Machine learning on insurance claims data provides a powerful tool to improve the efficiency and affordability of plans and care offered to patients enrolled in employer-sponsored health plans. With more accurate rate-setting, health insurance companies can design nuanced plan attributes, reducing the cost of care for their members. Our ML model achieved 20\% improved accuracy in absolute predictive performance over traditional actuarial methods and was able to identify over 80\% of new concession opportunities available to Delphi. This allows underwriters to better price and retain $<$500 employer group customers. This study can be used by payers to give underwriters improved pricing guidance, retaining business and giving a better and more affordable experience to members.

\section*{Acknowledgements}
We thank our counterparts at Delphi for collaboratively working with us to validate our model against a production-grade model with real data.  Additionally, we thank the following people for their contributions: Kim Branson, Dilawar Syed, Laika Kayani, Wil Yu, Shahab Hassani, Alexandra Pettet, Derek Gordon, Ash Damle, Thomas Watson,  Leon Barovic, Diana Rypkema and our investors at Blue Cross Blue Shield Venture Fund, and Khosla Ventures.

\bibliography{bibliography.bib}
\appendix
\newpage
\section{Appendix}

\subsection{Data Quality}\label{sec:a3}

The following describes some implications of modeling individual-level cost when the data model, from which the data is derived, considers the employer group as the primary entity.

\begin{itemize}
\item The cost of a claim depends entirely on what date you are inspecting the claim. The example in Figure \ref{fig:figDQ} shows hows, depending on the time period, a claim which has been reversed has a material impact on the total allowed amount we calculate. A corollary is that calculations done to a claims dataset which has been filtered on a date (so that all the claims occur prior to that date) can change as data after the filter date is added back in.
\end{itemize}

\begin{itemize}
\item Must consider ``alteration of training set" implications due to modeling at group-level. Members can move between groups because they change jobs or move to their partner's health plan. This raises the potential of a member being in both the training and testing data. Of course, the solution here, as above, is to remember that enrollment is only relevant with respect to a particular date, since enrollment can change.
\end{itemize}

\begin{itemize}
\item Finally, a rigorous accounting of the (1) count of member, (2) count of groups, and (3) total cost per group, from the insurance carrier's side, 
to receipt of raw data, to generating cost predictions from the pipeline, and finally to sending predictions back to the insurance carrier, is crucial to 
ensure the credibility of the performance estimates asserted by the model training process.
\end{itemize}

\renewcommand{\thefigure}{A1}

\begin{figure}[!b]
\centering
  \includegraphics[width=8cm]{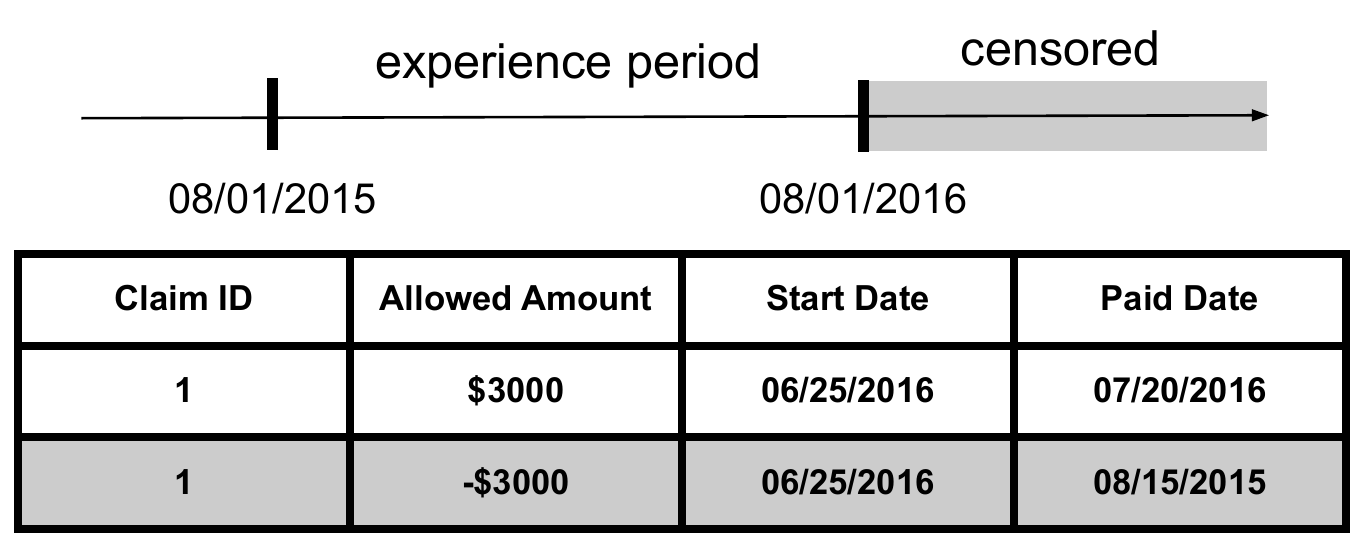}
  \caption{An example of a reversed claim that is seen differently due to censored data. In this example, the blackout period begins on 08/01/2016, so 
  all the claims with paid dates on or after this date are removed before we received the data. In reality, claim ``1" was originally incurred and paid before 08/01/2016 but was reversed on a paid date after 08/01/2016. However, because the data were censored based on paid dates, we saw that 
  claim ``1" had an allowed amount of \$3000. Once we receive the uncensored full claims data, we would see that claim ``1" had 
  actually been reversed and therefore cost \$0.}
  \label{fig:figDQ}
\end{figure}

\newpage
\subsection{Lumiata Data Model}\label{sec:ldm}
The Lumiata Data Model (LDM) is the standard format in which all Lumiata client data is formatted.  The purpose of this format is to consolidate all relevant patient information into a single place, with one record per patient. The schema of LDM is shown below:
\begin{verbatim}
root
 |-- MBR: string (nullable = true)
 |-- patient: struct (nullable = true)
 |    |-- patient_id: string (nullable = true)
 |    |-- birthday: date (nullable = true)
 |    |-- gender: string (nullable = true)
 |    |-- marital_status: string (nullable = true)
 |    |-- postal_code: string (nullable = true)
 |    |-- dead: boolean (nullable = true)
 |    |-- extensions: array (nullable = true)
 |    |    |-- element: struct (containsNull = true)
 |    |    |    |-- url: string (nullable = true)
 |    |    |    |-- valueString: string (nullable = true)
 |    |    |    |-- valueDate: date (nullable = true)
 |    |    |    |-- valueDecimal: float (nullable = true)
 |    |-- population_type: string (nullable = true)
 |-- practitioners: array (nullable = true)
 |    |-- element: struct (containsNull = true)
 |    |    |-- resource_id: string (nullable = true)
 |    |    |-- npi: string (nullable = true)
 |    |    |-- specialty_codes: array (nullable = true)
 |    |    |    |-- element: string (containsNull = true)
 |-- terms: array (nullable = true)
 |    |-- element: struct (containsNull = true)
 |    |    |-- resource_id: string (nullable = true)
 |    |    |-- date: date (nullable = true)
 |    |    |-- resource_type: string (nullable = true)
 |    |    |-- code: struct (nullable = true)
 |    |    |    |-- code: string (nullable = true)
 |    |    |    |-- system: string (nullable = true)
 |    |    |-- day_supply: integer (nullable = true)
 |    |    |-- display: string (nullable = true)
 |    |    |-- value: string (nullable = true)
 |    |    |-- interpretation: string (nullable = true)
 |    |    |-- units: string (nullable = true)
 |    |    |-- original_code: string (nullable = true)
 |    |    |-- practitioners: array (nullable = true)
 |    |    |    |-- element: string (containsNull = true)
 |    |    |-- resource_id_list: array (nullable = true)
 |    |    |    |-- element: string (containsNull = true)
 |    |    |-- lab_info: struct (nullable = true)
 |    |    |    |-- actual_value: string (nullable = true)
 |    |    |    |-- description: string (nullable = true)
 |    |    |    |-- comparator: string (nullable = true)
 |    |    |    |-- reference_range: string (nullable = true)
 |    |    |    |-- low_val: float (nullable = true)
 |    |    |    |-- high_val: float (nullable = true)
 |    |    |    |-- abnormal_result_description: string 
                    (nullable = true)
 |-- claims: array (nullable = true)
 |    |-- element: struct (containsNull = true)
 |    |    |-- resource_id: string (nullable = true)
 |    |    |-- paid_date: date (nullable = true)
 |    |    |-- start_date: date (nullable = true)
 |    |    |-- end_date: date (nullable = true)
 |    |    |-- fill_date: date (nullable = true)
 |    |    |-- billed_amount: float (nullable = true)
 |    |    |-- allowed_amount: float (nullable = true)
 |    |    |-- member_cost_share: float (nullable = true)
 |    |    |-- cob_amount: float (nullable = true)
 |    |    |-- paid_amount: float (nullable = true)
 |    |    |-- capitation_type: string (nullable = true)
 |    |    |-- major_service_category: string (nullable = true)
 |    |    |-- service_major: string (nullable = true)
 |    |    |-- coverage: string (nullable = true)
 |    |    |-- place_type: string (nullable = true)
 |    |    |-- surcharge_amount: float (nullable = true)
 |    |    |-- terms: array (nullable = true)
 |    |    |    |-- element: struct (containsNull = true)
 |    |    |    |    |-- resource_id: string (nullable = true)
 |    |    |    |    |-- date: date (nullable = true)
 |    |    |    |    |-- resource_type: string (nullable = true)
 |    |    |    |    |-- code: struct (nullable = true)
 |    |    |    |    |    |-- code: string (nullable = true)
 |    |    |    |    |    |-- system: string (nullable = true)
 |    |    |    |    |-- day_supply: integer (nullable = true)
 |    |    |    |    |-- display: string (nullable = true)
 |    |    |    |    |-- value: string (nullable = true)
 |    |    |    |    |-- interpretation: string (nullable = true)
 |    |    |    |    |-- units: string (nullable = true)
 |    |    |    |    |-- original_code: string (nullable = true)
 |    |    |    |    |-- practitioners: array (nullable = true)
 |    |    |    |    |    |-- element: string (containsNull = true)
 |    |    |    |    |-- resource_id_list: array (nullable = true)
 |    |    |    |    |    |-- element: string (containsNull = true)
 |    |    |    |    |-- lab_info: struct (nullable = true)
 |    |    |    |    |    |-- actual_value: string 
                              (nullable = true)
 |    |    |    |    |    |-- description: string (nullable = true)
 |    |    |    |    |    |-- comparator: string (nullable = true)
 |    |    |    |    |    |-- reference_range: string 
                              (nullable = true)
 |    |    |    |    |    |-- low_val: float (nullable = true)
 |    |    |    |    |    |-- high_val: float (nullable = true)
 |    |    |    |    |    |-- abnormal_result_description: string
                              (nullable = true)
 |    |    |-- prescription_written_date: date (nullable = true)
 |    |    |-- group_number: string (nullable = true)
 |    |    |-- specialty_codes: string (nullable = true)
 |    |    |-- revenue_codes: string (nullable = true)
 |    |    |-- is_capitation: boolean (nullable = true)
 |    |    |-- claim_id: string (nullable = true)
 |-- coverage: array (nullable = true)
 |    |-- element: struct (containsNull = true)
 |    |    |-- resource_id: string (nullable = true)
 |    |    |-- group_number: string (nullable = true)
 |    |    |-- branch_code: string (nullable = true)
 |    |    |-- start_date: date (nullable = true)
 |    |    |-- end_date: date (nullable = true)
 |    |    |-- account_effective_date: date (nullable = true)
 |    |    |-- account_termination_date: date (nullable = true)
 |    |    |-- medical_plan_type: string (nullable = true)
 |    |    |-- beneficiary_id: string (nullable = true)
 |    |    |-- account_state: string (nullable = true)
 |    |    |-- account_coverage_type: string (nullable = true)
 |    |    |-- account_renewal_month: string (nullable = true)
 |    |    |-- account_name: string (nullable = true)
 |    |    |-- scp_copay: float (nullable = true)
 |    |    |-- pcp_copay: float (nullable = true)
 |    |    |-- pcp_coinsurance: float (nullable = true)
 |    |    |-- scp_coinsurance: float (nullable = true)
 |    |    |-- individual_deductible: float (nullable = true)
 |    |    |-- collective_deductible_indicator: string 
              (nullable = true)
 |    |    |-- inpatient_coinsurance: float (nullable = true)
 |    |    |-- outpatient_coinsurance: float (nullable = true)
 |    |    |-- individual_oop_maximum: float (nullable = true)
 |    |    |-- medical_member_cost_share_percentage: float 
              (nullable = true)
 |    |    |-- pharmacy_member_cost_share_percentage: float 
                (nullable = true)
 |    |    |-- total_member_cost_share_percentage: float 
              (nullable = true)
 |    |    |-- product_type: string (nullable = true)
 |    |    |-- relationship: string (nullable = true)
 |    |    |-- industry_code: string (nullable = true)
 |    |    |-- pharmacy_coverage_indicator: string 
              (nullable = true)
 |    |    |-- effective_date: date (nullable = true)
 |    |    |-- cancel_date: date (nullable = true)
\end{verbatim}

\end{document}